\newcommand{\la}{\langle}
\newcommand{\ra}{\rangle}
\renewcommand{\d}{\partial}
\newcommand{\beq}{\begin{eqnarray}}
\newcommand{\eeq}{\end{eqnarray}}
\newcommand{\sbeq}{\begin{subeqnarray}}
\newcommand{\seeq}{\end{subeqnarray}}
\newcommand{\do}{^{\circ }}

\newcommand{\btem}{\bibitem}

\newcommand{\brho}{\mbox{{\boldmath $\rho$}}}
\newcommand{\balpha}{\mbox{{\boldmath $\alpha$}}}
\newcommand{\bbeta}{\mbox{{\boldmath $\beta$}}}

\newcommand{\btau}{\mbox{{\boldmath $\tau$}}}
\newcommand{\bpi}{\mbox{{\boldmath $\pi$}}}
\newcommand{\bsigma}{\mbox{{\boldmath $\sigma$}}}
\newcommand{\e}{{\rm e}}

\documentstyle[medium98,12pt,epsf]{article}
\begin{document}                        
\title{
Chiral Transition in Nuclear Medium and Relevant Observables}
\author{
T.~Kunihiro
       }
\address{Faculty of Science and Technology, Ryukoku University, 
 Seta, Otsu 520-2194, Japan}

\maketitle                              

\abstracts{
We discuss how the possible changes of hadron properties 
 can be tested, associated with the chiral
 transition in nuclear medium; the hadrons discussed are the vector mesons,
   the $\sigma$ meson and the chiral partner of the nucleon.
We  emphasize that the proper quantity is the spectral function 
 which describes the nuclear medium having a hadron; the analysis of
electro-quasielastic reactions, for instance, has a relevance to 
the physics of CERES.  
 The similarity  is indicated between 
  the physics discussed here with that of giant resonances.
Experiments using nuclear targets are proposed  
to observe the $\sigma$ meson and to  explore how the partial chiral 
restoration in nuclear medium manifests itself.
}

\section{Introduction}

Various theoretical approaches suggest that the QCD vacuum undergoes the 
phase transition with the order parameter $\la \bar{q}q\ra$ at
 high baryonic density $\rho_B$ as well as at high temperature 
 $T$\cite{ykis97,HK94,BR96}. However, our theoretical understanding on the 
 nature of the chiral transition at $\rho_B$ is relatively limited, although some suggestive results 
 have been obtained.
In other words, there are a lot of things to be
clarified and explored on the problem of the chiral transition at
$\rho_B$.

To have an idea on relevant observables to explore the problem,
let us start with recalling  the very fact that
 the chiral transition is a phase transition of the QCD vacuum.
In condensed matter physics and also in nuclear physics, a change of the
ground state (the vacuum) is examined  through the study of a possible
change of collective excitations on top of the ground states\cite{GK,HK94}:
 If the phase
transition is of second order or of weak first order, there will appear
characteristic changes in some specific collective modes, which are
actually the fluctuations of the order parameter or modes coupled to them.
Collective excitations in QCD are hadrons, 
especially the low lying ones. Therefore,
the study of hadrons in the nuclear medium has the  possibility 
to give the insight into the nature of the chiral restoration at 
finite density.  Indeed, there are suggestions on how some hadrons
change their properties in association with changes of the QCD vacuum
\cite{HK94,BR96}.

When a hadrons is put in a nucleus,
the hadron will couple strongly to various excitations in the system, 
such as nuclear particle-hole (p-h)  and  $\Delta$-hole excitations, 
simultaneous excitations of them and mesons and so on:
In general, the hadron may dissociate into complicated excitation to loose its 
dentity in the nuclear medium. 
The relevant quantity is the response function or spectral function 
of the system when the quantum numbers of the hadron are put in.
If the coupling of the hadron  with the environment is relatively small,
then there may remain a peak with a small width in the spectral function, 
which  correspond to the hadron; such a peak 
may be viewed as an elementary excitation or a quasi particle known in 
Landau's Fermi liquid theory for fermions. 
Landau gave an argument that there will be a chance to describe a system as 
an assembly of almost  free quasi-particles owing to the Pauli principle 
  when the temperature is low. 
 Symmetries of the system may ensure 
  the existence of an elementary bosonic excitation in the many-body 
  system\cite{GK,HK85,su}.

In this report, we discuss three topics on possible changes of hadron 
properties in the nuclear medium; the hadrons to be discussed are 
the $\rho$ meson, the sigma meson and the possible chiral partner of the
 nucleon. 
 
\section{The spectral function in the vector channel and the inclusive
electron  scattering}

Various effective models, the QCD sum rules and the scaling argument
all predicted a decrease of the vector meson masses $m_{_V}$.
The subsequent measurement of the lepton pairs from the heavy-ion 
collisions by CERES group \cite{ceres} seemed to show a decrease of the 
spectral function in the $\rho$ meson channel, which might be 
an evidence of the decrease of $m_{_V}$\cite{friman}.
Thus the experiment prompted heated discussions on possible interpretation
 of the data.
 
The cross section in this reaction is essentially the spectral function
\beq
R_{L, T}(\omega, q)=\frac{\sigma(\e ^{+}\e^{-}\rightarrow {\rm hadrons})}
          {\sigma(\e ^{+}\e^{-}\rightarrow \mu^{+}\mu^{-})}
     \propto {\rm Im}\Pi_{L, T}(\omega, q),
\eeq
where $\Pi_{L,(T)}(\omega, q)$ is  the longitudinal (transverse) response
function in the $\rho$ meson channel; the spectral function is given by the
 imaginary part of the response function or the retarded Green's function.
At $\rho_B\not=0$,  $\Pi_{L}(\omega, q)\not=\Pi_{T}(\omega, q)$ when 
$q\not=0$, and manybody effects as well as the lack of the Lorenz
 invariance gives rise to a significant modification of the 
 spectral function from that in the free space.
 For example, the coupling of the process
$\rho \rightarrow   {\rm p-h}\ (\Delta{\rm -h}) + \pi$,which makes a 
collective mode with the $\rho$ meson quantum number,
increases the width of the vector meson. Furthermore the $p$-wave coupling
of the $\rho$ meson with baryons induces the momentum dependence of the
spectral function. Indeed there exist many baryons, i.e., excited states of
the nucleon and the $\Delta$, which couple to the $\rho$ meson; the 
$(\omega, q)$-regions of some of the baryon-hole excitations overlap
with the dispersion relation of $\omega =\sqrt{m_{\rho}^2+q^2}$,
which makes the spectral function in the $\rho$ meson channel complicated;
see Fig.1, where the $(\omega,q)$-regions of various baryon-hole
excitations are shown together with the $\rho$ meson dispersion 
relation (in the free space).

\begin{figure}[bht] 
 \begin{center}
   \input{disp_fig}
 \end{center}
\label{fig.1}
\end{figure}
\setcounter{figure}{1}

However, there are some ambiguities in the theoretical calculations
based on such a hadron scenario; many coupling constants of the vector 
meson with baryons including their form factors are obviously 
not available.
In fact, S.H. Lee has shown that the QCD sum rule approach gives different 
momentum dependence of the dispersion relation of the $\rho$ meson\cite{shl}.

The problem is to  extract the global structure of the
 spectral function in the $\rho$ channel and understand it theoretically.
The moral extracted from these studies may be that putting a hadron
in a nucleus is nothing but providing the same quantum numbers 
 as those possessed by the hadron. And the proper observables for this
 physics is the response function or the spectral function in this
  channel, which might show peaks corresponding to the hadron as
   an elementary excitation in the system.
   
The reader should have recognized that the underlying physics of
 this game is the same as that of giant resonances\cite{gr}: The physics of
  the giant resonances are actually concerned with the global 
  structure of the response function in the relevant channel.
The analogy or correspondence goes further:
Spin-isospin dependent giant resonances are related with the
 pion and the $\rho$ meson in nuclei\cite{kuni81,suzuki}:
The possible softening of the longitudinal
spin-isospin mode in the large momentum region (or 
modes with a large angular momentum)
is a precursor of the pion condensation, 
 a phase transition of nuclear matter\cite{kuni81}. 
 This is an example for the
 general feature that there arises a soft mode as a precritical
  phenomena of the phase transition, as mentioned before.

We notice that exploring the spectral function (the response function)
in the $\rho$ channel  has a relevance to  a longstanding problem in 
the inclusive electron scattering\cite{moniz}:
 The cross section of the electro-quasi-elastic 
reaction  in the transverse channel has an anomalously large cross 
section in the energy region 
between the nucleon- and the $\Delta$-production region. The origin of 
the anomalous strength may be attributed to a possible $\Delta$-hole
 collective excitation due to the strong $p$-wave $\rho$-N$\Delta$ 
 tensor coupling\cite{kuni82}; the collective mode supplies a strength
 in the $(\omega,q)$ region below the free $\Delta$-hole dispersion 
curve, which region is found to be kinematically hit 
by the quasi-free electron scattering. 
 Indeed, the {\em attractive} $\rho$-baryon tensor coupling 
 $B^{\dagger}\bsigma \btau B'\times \nabla \brho$ in the transverse channel 
could make the $\rho$-meson 
  condensation in nuclear matter at high density\cite{kuni78}.

\section{The sigma meson and the spectral function in the sigma channel
 in the nuclear medium}

 The sigma meson is the chiral partner of the pion for the 
$SU_L(2)\otimes SU_R(2)$ chiral symmetry in QCD.
The   particle representing the quantum fluctuation of the order parameter
 $\tilde {\sigma}\sim \la (:\bar q q:)^2\ra$ is the sigma meson. 
The sigma meson is analogous
 to the Higgs particle in the Weinberg-Salam theory.
Some effective theories and Weinberg's mended symmetry 
predict the sigma meson mass $m_{\sigma}\sim 500$-700 MeV.
Such a  scalar meson with a low mass can account for various
 experimental and empirical facts\cite{HK94,kuni95}.

A tricky point on the sigma meson is that the sigma meson strongly couples 
to two pions which  gives rise to a huge width $\Gamma \sim m_{\sigma}$.
Recent phase shift analyses of the pi-pi scattering in the scalar channel
 claim a pole of  the scattering matrix in the complex energy plane  
 with the real part Re$m_{\sigma}= 500$-700 MeV and the imaginary part
 Im$m_{\sigma}\simeq 500$MeV\cite{pipi}.

Since the sigma meson is the fluctuation of the order parameter of the 
chiral transition, it will become a soft mode 
and induce characteristic phenomena associated with the chiral 
restoration in a hot and/or dense nuclear medium. 
Thus one may expect a better chance to see the sigma meson in a 
clearer way in a hot and/or dense medium than in the vacuum\cite{HK94}.

The present author proposed three types of experiment to produce the sigma 
meson in a nucleus\cite{kuni95}; the probes are
 pions, protons, light nuclei and  electrons.
 To detect the sigma, one may use 4 $\gamma$'s and/or two leptons. The 
 latter
 process is  possible when the $\sigma$ has a three momentum
 owing to the scalar-vector couping at $\rho_B\not=0$.

 In the introduction, we emphasized that the relevant 
 observable is the spectral function.
Recently, a calculation of the spectral function in the sigma channel
has been performed with the $\sigma$-2$\pi$ coupling incorporated 
in the linear $\sigma$ model at finite $T$; it was shown that 
the enhancement of the spectral function in the $\sigma$-channel
just above the
two-pion threshold is the most distinct signal of the softening \cite{CH98}.

More recently, it has been shown \cite{HKS} that the spectral enhancement
associated with the partial chiral restoration  
takes place also at finite baryon density close to 
$\rho_0 = 0.17 {\rm fm}^{-3}$.
Refering to \cite{HKS} for the  explicit model-calculation, let us
 describe the general
 features of the spectral enhancement near the two-pion threshold.
 Consider the propagator 
 of the $\sigma$-meson at rest in the medium :
$D^{-1}_{\sigma} (\omega)= \omega^2 - m_{\sigma}^2 - $
$\Sigma_{\sigma}(\omega;\rho)$,
where $m_{\sigma}$ is the mass of $\sigma$ in the tree-level, and
$\Sigma_{\sigma}(\omega;\rho)$ is 
the loop corrections
 in the vacuum as well as in the medium.
 The corresponding spectral function is given by 
\beq
\rho_{\sigma}(\omega) = - \pi^{-1} {\rm Im} D_{\sigma}(\omega).
\eeq

We parametrize the chiral condensate in nuclear matter
 $\langle \sigma \rangle$ as
\beq
\label{cond}
\langle \sigma \rangle \equiv  \sigma_0 \ \Phi(\rho).
\eeq
In the linear density approximation,
 $\Phi(\rho) = 1 - C \rho / \rho_0 $ with
 $C = (g_{\rm s} /\sigma_0 m_{\sigma}^2) \rho_0$.
 Instead of using $g_{\rm s}$, we  use $\Phi$  as a basic parameter in the 
 following analysis.  The plausible value of $\Phi(\rho = \rho_0)$ is
 0.7 $\sim$ 0.9 \cite{HK94}.

\begin{figure}[tbh]
\begin{center}
\epsfile{file=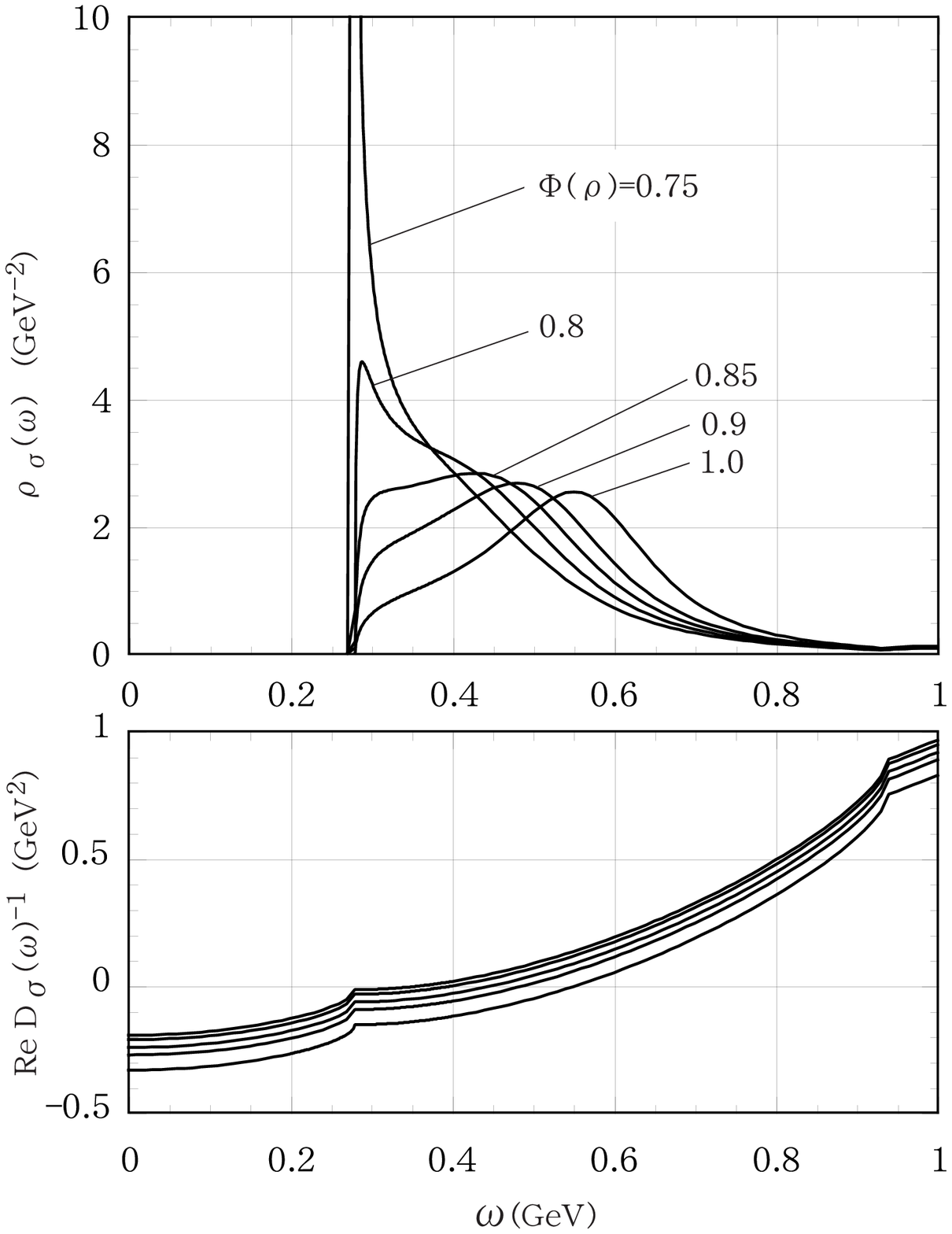,width=0.50\textwidth}
\end{center}
\caption{Spectral function for $\sigma$ and  the 
 real part of the inverse propagator for several values of
 $\Phi = \la \sigma \ra / \sigma_0$ with
 $m_{\sigma}^{peak} = 550$ MeV. In the lower panel,
 $\Phi$ increases from bottom to top.}
\label{fig.2}
\end{figure}
 
The spectral function together with ${\rm Re} D_{\sigma}^{-1}(\omega)$  
 calculated with a linear sigma model are shown 
  in Fig.2: The characteristic enhancements of the spectral
 function just above the 2$m_{\pi}$.
The mechanism of the enhancement is understood as follows.
The partial restoration
 of chiral symmetry implies that  $m_{\sigma}^*$ approaches toward
 $m_{\pi}$. On the other hand,
 ${\rm Re}D^{-1}(\omega)$ has a cusp at $\omega = 2 m_{\pi}$.
The cusp point goes up with the density and
eventually hits the real axis at $\rho = \rho_c$
 because ${\rm Re}D^{-1}(\omega )$ increases associated
 with $m_{\sigma}^* \rightarrow 2 m_{\pi}$.
It is also to be noted that even before
the $\sigma$-meson mass $m_{\sigma}^*$ and $m_{\pi}$ in the medium 
are degenerate,i.e., the chiral-restoring point, 
 a large enhancement
 of the spectral function near the $2m_{\pi}$ is seen.

To confirm the threshold enhancement,
measuring 2$\pi^0$ and 
$2\gamma$ in experiments with hadron/photon beams off
 the  heavy nuclear targets are useful. 
 Measuring $\sigma \rightarrow 2 \pi^0 \rightarrow
  4\gamma$ is experimentally feasible 
 \cite{4gamma}, which is free from the $\rho$ meson meson background
  inherent in the $\pi^+\pi^-$ measurement.
 Measuring of 2 $\gamma$'s from the electromagnetic decay of the $\sigma$
 is interesting because of the small final state
 interactions, although the branching ratio is small.\footnote{
One needs also to fight with large 
 background of photons mainly coming from $\pi^0$s.}
 Nevertheless,  if the enhancement is prominent,
  there is a chance to find the signal.  
 When $\sigma$ has a finite three momentum,
one can detect dileptons  
 through the scalar-vector mixing in matter: $\sigma \to \gamma^* \to
 e^+ e^-$. 

We remark that (d, $^3$He)  reactions is also useful to produce 
 the  excitations in the $\sigma$ channel in a nucleus because of the
large incident flux, as  the $\eta$ production\cite{HHG}.
 The incident kinetic energy $E$ of the  deuteron in the laboratory
 system is  estimated to be
  $1.1 {\rm GeV} < E < 10$ GeV, 
  to cover the spectral function 
 in the range  $2m_{\pi} < \omega < 750$ MeV.

Recently  CHAOS collaboration  \cite{ppp} measured the 
$\pi^{+}\pi^{\pm}$
invariant mass distribution $M^A_{\pi^{+}\pi^{\pm}}$ in the
 reaction $A(\pi^+, \pi^{+}\pi^{\pm})X$ with the 
 mass number $A$ ranging
 from 2 to 208: They observed that
the   yield for  $M^A_{\pi^{+}\pi^{-}}$ 
 near the 2$m_{\pi}$ threshold is close to zero 
for $A=2$, but increases dramatically with increasing $A$. They
identified that the $\pi^{+}\pi^{-}$ pairs in this range of
 $M^A_{\pi^{+}\pi^{-}}$ is in the $I=J=0$ state.
The $A$ dependence of the 
 the invariant mass distribution presented in \cite{ppp} 
 near 2$m_{\pi}$ threshold has a close
 resemblance to our model calculation in Fig.2, which suggests
 that this experiment may already provide
  a hint about how the partial restoration of chiral symmetry
 manifest itself at finite density.\footnote{See \cite{wambach} for
 other approaches to explain the CHAOS data.}

\section{Parity doubling of baryons}
So far we have discussed the meson properties in the nuclear medium 
in relation with the possible chiral restoration.
Then, do baryons show up any characteristic features when the chiral
symmetry is (partially) restored? To answer this question, one needs
 to clarify how baryons can be described in the context of chiral
symmetry.
There are actually some approaches on this problem; for
 example, we know the nonlinear representation theory and the Skyrm
 model of baryons. Here, we introduce a linear sigma model where
  baryons are described in the linear representation
  of the chiral symmetry, and discuss its
  phenomenological consequences.
   
About a decade ago, motivated by the lattice simulation on the
screening masses of the nucleons with positive and negative parity
both in chirally broken and restored phases,
De Tar and the present author proposed a linear sigma model with 
parity doubling\cite{detar}. The model reads,
\beq
{\cal L}&=&\bar{\Psi}i\gamma\cdot \d \Psi- g_1\bar{\Psi}(\sigma+
         i{\btau}\cdot{\pi}\rho_3\gamma_5)\Psi, \nonumber \\  
 \ \    & & +g_2\bar{\Psi}(\rho_3\sigma +i{\btau}\cdot{\bpi}\gamma_5)
             \Psi -iM_0\bar{\Psi}\rho_2\gamma_5\Psi\nonumber \\ 
        & & +{\cal L}_M(\sigma, {\bpi}),
\eeq
with $\Psi=\ ^t(\psi_1, \psi_2)$
where $\psi_1$ and $\psi_2$ are both a Dirac spinor and $\rho_i$
 ($i=1, 2, 3$) is the Pauli matrices acting on the spinors composed of
  $\psi_1$ and $\psi_2$.  $\psi_i$ ($i=1, 2$) are
supposed to be the ur-state of the positive-parity and the negative-parity
 nucleon, respectively; the term with $\rho_2$ mixes the two 
Dirac spinors to make the physical eigenstates $\Psi_{+}$ and $\Psi_{-}$
with positive and negative parity, respectively.
 
This lagrangian can be shown invariant under the extended chiral 
transformation,
\beq
\delta \Psi=1/2\cdot i {\balpha}\cdot{\btau}\Psi+ 
           1/2\cdot i {\bbeta}\cdot{\btau}\rho_3\gamma_5\Psi.
\eeq            

In the tree level, the eigen-value problem is solved yielding,\\ 
$\Psi_{+}(p)=N\ ^t(\psi _{+}(p),\e^{-\theta}\gamma_5\psi_{+}(p))$,
$\Psi_{-}(p)=N\ ^t(-\e^{-\theta}\gamma_5\psi_{-}(p),\psi_{-}(p))$,
with
\beq
\gamma\cdot p \psi_{\pm}(p)=M_{\pm}\psi_{\pm}(p),
\quad 
M_{\pm}=\mp g_2\sigma_0+\sqrt{(g\sigma_0)^2+M_0^2}
\label{eq:mass},
\eeq
where $N=1/\sqrt{1+{\rm exp}(-2\theta)}$, which is determined by 
the normalization conditions $\Psi_{\pm}\Psi_{\pm}=1$.
We remark that (1) $M_{-}>M_{+}$ and 
(2) $M_{\pm}\rightarrow \vert M_0\vert \not=0$.

It can be shown that   
 ${\Psi}_{-}$ has the opposite parity and the charge 
 conjugation to $\Psi_{+}$.
 
The axial charge matrix is calculated to be 
\beq
\hat{g}_A=\pmatrix{\tanh \theta \ \quad -1/\cosh \theta \cr
                   -1/\cosh \theta \ -\tanh \theta
                   },
\eeq
where the 1-1 (2-2) component denotes 
$(g_A)_{NN}$ ($(g_A)_{N'N'}$), for example.  We remark that the
negative parity nucleon has a negative axial charge in this model.
One can also show that the generalized Goldberger-Treiman relation;
\beq
g_{\pi NN}=g_{ANN}M_{+}/\sigma _0,\quad 
g_{\pi N'N'}=g_{AN'N'}M_{-}/\sigma _0,\quad 
g_{\pi NN'}=g_{ANN'}(M_{+}-M_{-})/2\sigma _0.
\eeq

So far, we have not identified the negative parity nucleon N' with 
any particle.  There are two possibilities; one is that 
 N'$\equiv$ N(1535) and the other is that N' is elusive like the sigma 
 meson and has not been observed experimentally, yet.
Here, let us assume the first possibility. Then, from the decay width
$\Gamma _{\pi N}\simeq 70$ MeV for N'$\rightarrow $N$+\pi$,
and the empirical values 
$M_{+}=939, \quad M_{-}=1535, \quad \sigma _0=93$,
in MeV, we have 
$g_{\pi NN'}=0.70$, which correspond to the values of the model 
parameters as follows;
$\sinh \theta =5.5,\ \  g_1=13.0, \ \ g_2=3.2, M_0=270\ ({\rm MeV})$.
\begin{figure}[hbt] 
 \begin{center}
   \input{m_baryon_fig}
 \end{center}
\end{figure}
How can the model be tested experimentally?
 A characteristic feature of this model is that the axial charge
 of the nucleon and N(1535) has the opposite sign, provided that 
 N' is identified with N(1535). Therefore experiments which can see an 
interference between the two amplitudes including
 $g_{\pi NN}$ and $g_{\pi N'N'}$, respectively, will be interesting.

It is also interesting to explore the properties of the negative-parity
 nucleon experimentally.
Assuming that the sigma condensate $\sigma _0$ decreases with 
the baryonic density, we can discuss some phenomenological
 consequences of the model at finite density: Fig.3 shows the 
 $\sigma_0$ dependence of the masses $M_{\pm}(\sigma_0)$. 
 One can see  that both the masses
$M_N$ and $M_{N'}$ decrease as the chiral symmetry is restored, which 
suggests that a downward mass shift of both the particles at finite 
density.  The rate of the decrease is larger for N(1535) than the nucleon.
The axial charges and the coupling with the pion of both the 
nucleon decrease as the chiral symmetry is restored,
 which may give another
origin of the quenching of the $\beta$ decay in nuclei and affect the
cooling rate of neutron stars; see \cite{jido} for recent developments.
 
\section{Summary}
This report may be summarized as follows:\
(1)
 The chiral transition (or a change of the QCD vacuum generically) may 
 imply that of properties  such as mass shifts and a
  change of the life time  of hadrons 
  as (collective) excitations on top of
  the vacuum.  The hadrons discussed in this context are the sigma meson, 
   the $\rho$ meson and the negative-parity nucleon.\
(2) 
If a hadron is put in in the nuclear medium, the  coupling of the hadron 
with the environment may be so large that it becomes inadequate to 
describe the system in terms of the mass and the width of the hadron.
The proper quantity to describe the situation properly is the strength 
function in the hadron channel. \
(3) Nevertheless, if the hadron exists as a quasi-particle not loosing its
identity, it is still adequate to interpret experimental data in terms of 
the mass and the width of the hadron.\
(4) Experiments using light ions and electrons as probes are as useful
to explore the physics discussed here as super-relativistic heavy ion 
collisions. They are also complementary to each other.\
(5) The anomalous enhancement near two-$m_{\pi}$ threshold in the 
strength function in $I=J=0$ channel (the sigma channel) is interesting to
see how the partial restoration of the chiral symmetry manifest itself 
in the nuclear medium. Some experiments have been proposed on this problem.

\acknowledgments
The author is grateful to the organizers of the workshop
 to give him a chance  to present a talk at 
the workshop held on the occasion of the 60-years birthday of 
Prof. Ichimura. 
A part of this report is based on the work done in collaboration with 
 T. Hatsuda and H. Shimizu. He acknowledges them for discussions and the 
collaboration. This work was  partly supported by   the Japanese
Grant-in-Aid for Science and Research Fund of the Ministry of Education,
 Science and Culture, No. 09640377 and No.08304024.

\end{document}